\documentclass[reprint,onecolumn,notitlepage,showpacs,preprintnumbers,amsmath,amssymb,aps,pra]{revtex4-1}

\usepackage[pdftex]{graphicx}
\usepackage{dcolumn}
\usepackage{bm}
\usepackage{psfrag}
\usepackage{footnpag}
\usepackage{amsmath,amssymb}
\usepackage{nccfoots}
\usepackage{stmaryrd}

\begin{document}

\preprint{arXiv/}

\title{Single-step Propagators for calculation of time evolution in quantum systems with arbitrary interactions}

\author{Ivan~Gonoskov}
\email{ivan.gonoskov@gmail.com}
\affiliation{Department of Physics, Ume\aa\ University, SE-90187 Ume\aa, Sweden}

\date{\today}

\author{Mattias~Marklund}
\email{mattias.marklund@physics.umu.se}
\affiliation{Department of Physics, Ume\aa\ University, SE-90187 Ume\aa, Sweden}
\affiliation{Department of Applied Physics, Chalmers University of Technology, SE-41296, Gothenberg, Sweden}

\begin{abstract}
We propose and develop a general method of numerical calculation of the wave function time evolution in a quantum system which is described by Hamiltonian of an arbitrary dimensionality and with arbitrary interactions. For this, we obtain a general n-order single-step propagator, which could be used for the numerical solving of the problem with any prescribed accuracy. We demonstrate an applicability of the proposed approach by considering a propagation of an electron in focused electromagnetic field with vortex electric field component.
\end{abstract}

\maketitle

\section{Introduction}

	Theoretical analysis of quantum systems and their time evolution often leads to necessity of solving the corresponding partial differential equations with time-dependent Hamiltonians. For the typical problem, when the corresponding closed-form solution is unknown, a direct numerical calculation could be an efficient tool for obtaining the solution approximation with a reasonable (and sometimes controlled) accuracy level. The main aim of our manuscript is to develop a general way of the numerical implementation with controlled accuracy based on single-step finite-order propagators. Such numerical schemes represent a step-by-step repetition procedure for a number of steps corresponding to some certain time intervals. At each step, a new solution approximation is calculated, based on the previous one (or initial condition), and a certain operator: single-step propagator.


	A derivation of the single-step propagators is usually based on a power series expansion of the corresponding solution approximation obtained from a perturbation theory or its analogues (see \cite{FF, SpO-Drozdov, Num-Drozdov, Be} and references therein). The particular propagators and the corresponding implementations further could be modified in a variety of ways, in order to increase the numerical efficiency for a certain problem. For example, for solving a time-dependent Scr\"oedinger equation (TDSE) one of the most popular and well-known method is an \textit{exponential split-operator technique} \cite{FF, JPhB}, which correspond to a second-order solution approximation in the case of separable Hamiltonians. Other improvements could be achieved by using special expansions and specific references for the general solution approximation (for example, \textit{harmonic-oscillator reference propagator} \cite{Osc-Drozdov, Num-Drozdov}). However, the majority of these methods could not be straightforwardly adapted and efficiently implemented for the most general statement which includes arbitrary Hamiltonian structures and interaction potentials in the corresponding equations. 

	In this manuscript we propose and develop a numerical approach which is based on general single-step propagators. Such propagators allow to calculate a time evolution in the case of a most general statement: for arbitrary Hamiltonian structures and interaction potentials, and with any prescribed accuracy level. The derivation of the general propagator is based on the exact solution from cyclic operator decomposition (COD) \cite{infCOD}. This solution is used for the obtaining of the $n$-order solution approximation for a finite time interval in a closed-form. Final expression for the propagator represents the optimization of the corresponding approximation for the efficient numerical implementation.
	
	The manuscript is organized as follows. After the introduction in Sec.I, we consider the exact solution, finite-order approximations and the corresponding propagators in Sec.II. In Sec.III we derive the general $n$-order single-step propagator. Next, in Sec.IV we consider several particular examples and corresponding optimizations for the \textit{first-}, \textit{second-}, and \textit{third-order} propagators. In Sec.V we apply the propagators technique for the calculation in the case of non-separable Hamiltonian: namely, the interaction of the electron in focused electromagnetic field with vortex electric field component. Sec.VI is devoted to conclusions.
\newpage	

\section{Exact solutions and finite-order propagators}	
	
	A time evolution of a quantum system could be described by the wave function which is a solution of the corresponding time-dependent partial differential equation \cite{Schiff}. In a general way such equation could be written as follows:
\begin{equation}\label{qe1}
i\hbar\,\partial_{t}\Psi=\hat{H}\Psi,
\end{equation} 
where $\hat{H}=\hat{H}(t,x_{1},...,x_{n},\hat{p}_{1},...,\hat{p}_{n})$ is a Hamiltonian including all the interactions, and $\Psi=\Psi(t,x_{1},...,x_{n},\hat{p}_{1},...,\hat{p}_{n})$ is an unknown wave function, which describes in some certain representation the time evolution of a quantum system. Here $x_{i}$ and $\hat{p}_{j}$ correspond to the canonical coordinates and momenta with the following commutation relations: $[x_{i},\hat{p}_{j}]=i\hbar\delta_{ij}$. Such formulation for some specific cases of the Hamiltonian definition and the corresponding wave function representation could lead, for example, to TDSE, Pauli or Dirac equations. For the sake of brevity, we will further omit the writing of the $x_{i}$ and $\hat{p}_{j}$ dependence in the corresponding operators and functions.

	In the case of the initial value problem (or Cauchy problem), the unknown solution $\Psi(t)$ is a result of the time evolution of the given initial condition $\Psi(t_{0})$ under the action of the Hamiltonian. Exact solution of the problem could be written in terms of exact propagator $\hat{P}$ for the finite time interval:

\begin{equation}\label{qe2}
\Psi(t)=\hat{P}(t_{0},t)\Psi(t_{0}),
\end{equation} 

	Since the Eq.(\ref{qe1}) is a linear equation, the solution and the corresponding exact propagator could be obtained by using the theory of cyclic operator decomposition (COD). In the case of the initial value problem, the generating function could be chosen as the initial condition $\Psi(t_{0})$, and the corresponding cyclic operators depend on Hamiltonian operator $\hat{H}(t)$ and time as follows (see \cite{infCOD} for full details): 

\begin{equation}\label{qe3}
\begin{aligned}
\Psi(t)=&\hat{P}(t_{0},t)\Psi(t_{0})=\left[1+\sum\limits_{m=1}^{\infty}\left(\hat{G}^{-1}\hat{V}\right)^{m}\right]\Psi(t_{0}) \,\,,  &&&&&&&&&&\\
&\text{where:}\\
&\hat{V}\phi(t)=\frac{1}{i\hbar}\hat{H}(t)\phi(t)\,,\,\,\,\, \forall{\phi(t)},\\
&\hat{G}^{-1}\varphi(t) = \int\limits_{t_{0}}^{t}\varphi(\tau)d\tau\,,\,\,\,\, \forall{\varphi(t)}.
\end{aligned}
\end{equation} 

	The convergence of such or similar series is analyzed in \cite{infCOD, Corbett}. Note, that the correctness of the solution Eq.(\ref{qe3}) could be checked by the direct substitution in Eq.(\ref{qe1}).

	The solution Eq.(\ref{qe3}) could be useful for a theoretical analysis, but it is not always convenient for the direct calculations because of the infinite series and $\hat{G}^{-1}$ operator. Therefore, various approximations of this solution which are applicable for the efficient numerical calculations, should be considered and analyzed. 

	For the efficient numerical calculation in the case of relatively small time step $\Delta{t}=t-t_{0}$ the finite-order propagator technique could be used. The order of propagator corresponds to the order of approximation of the exact solution, and it is equal to the highest power of $\Delta{t}$ in the series expansion of the operator. In this case the approximation is following:
	
\begin{equation}\label{qe4}
\Psi(t)\approx\hat{P}_{n}(t_{0},t)\Psi(t_{0}) + O(\Delta{t}^{n+1}),
\end{equation}  
where $\hat{P}_{n}(t_{0},t)$ is the $n$-order propagator, and the accuracy level is determined by the absolute error, which is not more than order of $\left|O(\Delta{t}^{n+1})\right|$.

	So, the goal of our manuscript is to obtain such $n$-order propagator, which could be useful for the efficient numerical implementation, without any specific restrictions for the Hamiltonian form, dimensionality and type of interactions. Since a time origin could be chosen arbitrary, below we assume for brevity: $t_{0}=0$, $\Delta{t} = t$.

-----------------------

\section{Derivation of n-order single-step propagator}

	For the numerical calculation based on single-step propagators technique it is necessary to obtain first the closed-form finite-order approximation of the corresponding exact solution. This means, that we need to take into account only finite number of terms from the Eq.(\ref{qe3}) and obtain appropriate approximation for the $\hat{G}^{-1}$ operator. For this we firstly consider the following rule of integration:
	
\begin{equation}\label{qe5}
\int\limits_{0}^{x}f(y)dy = \left[xf(x)\right] - \int\limits_{0}^{x}yf'(y)dy.
\end{equation}	

	The last integral in the Eq.(\ref{qe5}) could be again transform in the same way. The sequential repetition gives us:
	
\begin{equation}\label{qe6}
\int\limits_{0}^{x}f(y)dy = xf(x) - \frac{x^{2}}{2}f'(x) + \frac{x^{3}}{3!}f''(x) - ... = \sum\limits_{k=1}^{N}\Big[(-1)^{k+1}\frac{x^{k}}{k!}\,f^{(k-1)}(x)\Big] + (-1)^{N}\int\limits_{0}^{x}f^{(N)}(y)\,\frac{y^{N}}{N!}\,dy\;,
\end{equation}	
where $f^{(k)}(y)=d^{k}f(y)/dy^{k}$ is the corresponding $k$-th derivative. The expression is correct only in the case when the corresponding series is exist, for example if we work with analytical functions from a class $\mathbb{C}^{N}$. In the case when we work with smooth functions (class $\mathbb{C}^{\infty}$), which satisfy the condition: $\exists{}C_{0},\,\exists{}B\,:\;\left|f^{(n)}(y)\right|\leq{C_{0}B^{n}}, \forall{y}\in{[0,x]}, \forall{n}\geq{0}$, the Eq.(\ref{qe6}) could be useful for the obtaining an arbitrary finite-order approximation of the solution Eq.(\ref{qe3}) with any prescribed accuracy. 

	Thus, from the Eq.(\ref{qe3}), Eq.(\ref{qe6}) and under necessary conditions for the convergence and accuracy, we could conclude the following two points. First, for the derivation of the $n$-order propagator it is sufficient to take into account only first $(n+1)$ terms in the corresponding series in Eq.(\ref{qe3}). Second, the $R$-order approximation $\hat{G}^{-1}_{R}$ of the $\hat{G}^{-1}$ operator could be expressed as a series, as follows:

\begin{equation}\label{qe7}
\hat{G}^{-1}_{R} = \Big[ t - \frac{t^{2}}{2}\frac{\partial}{\partial{t}} + \frac{t^{3}}{3!}\frac{\partial^{2}}{\partial{t}^{2}} - ... + (-1)^{R+1}\frac{t^{R}}{R!}\frac{\partial^{R-1}}{\partial{t}^{R-1}}\Big] = \Big[ \sum\limits_{k=1}^{R}(-1)^{k+1}\frac{t^{k}}{k!}\hat{G}^{k-1} \Big]\,,
\end{equation}
where $\hat{G} = \partial_{t}$, ($\hat{G}^{0} \equiv 1$) is a \textit{left inverse operator} for the operator $\hat{G}^{-1}$:

\begin{equation}\label{qe8}
\begin{aligned}
&\hat{G}\phi(t)=\frac{\partial{\phi(t)}}{\partial{t}}\,,\,\,\,\forall{\phi(t)},\\
&\hat{G}\hat{G}^{-1} \equiv 1\,.
\end{aligned}
\end{equation}

	Note, that on the contrary to the $\hat{G}^{-1}$ operator, the $\hat{G}$ is equal to elementary operation: derivation, which could be calculated in closed-form for a given closed-form function. 

	For the derivation of the $n$-order approximation of the Eq.(\ref{qe3}) it is necessary to use different $R$-order approximations ($R(n)\leq{n}$) of the operator $\hat{G}^{-1}$ in each corresponding therm in the series. According to Eqs.(\ref{qe3}, \ref{qe7}) we could write: 
	
\begin{equation}\label{qe9}
\Big[\left(\hat{G}^{-1}\hat{V}\right)^{m}\Big]_{n} = \prod\limits_{p=1}^{m} \left(\hat{G}^{-1}_{R(n,p)}\hat{V}\right),
\end{equation}
where $R(n,p) = n-p+1$, and we use standard \textit{left-to-right} order of operator multiplication: $\prod\limits_{j=1}^{L}\hat{A}_{j} = \hat{A}_{1}\hat{A}_{2}\cdots\hat{A}_{L-1}\hat{A}_{L}$.

	Finally, we obtain the closed-form expression for the $n$-order propagator:
	
\begin{equation}\label{qe10}
\hat{P}_{n}(0,t)=1+\sum\limits_{m=1}^{n}\prod\limits_{p=1}^{m}\bigg\{ \Big[\sum\limits_{k=1}^{n-p+1}(-1)^{k+1}\frac{t^{k}}{k!}\hat{G}^{k-1} \Big]\hat{V} \bigg\}\;.
\end{equation}

	This expression includes all the necessary therms up to $n$-th order, and consists of the finite number of elementary operations (we suppose that the Hamiltonian includes only elementary operations).

-----------------------
	
\newpage
\section{Examples of finite-order propagators}

	In this section we consider some of the examples of the finite-order propagators, based on Eq.(\ref{qe10}). Let us start from the \underline{\textit{first-order propagator}}:
	
\begin{equation}\label{qe11}
\hat{P}_{1}(0,t)=1+\frac{t}{i\hbar}\hat{H}(t)\;.
\end{equation}	
	
	This expression could be used only for very approximate estimations, but not generally convenient for the numerical implementations due to the low accuracy; the error in each step is proportional to $O(t^{2})$.
	
	More interesting case is the \underline{\textit{second-order propagator}}:

\begin{equation}\label{qe12}
\hat{P}_{2}(0,t)=1 + \frac{t}{i\hbar}\hat{H}(t)-\frac{t^{2}}{2i\hbar}\partial_{t}\hat{H}(t) + \frac{t^{2}}{2(i\hbar)^{2}}\hat{H}^{2}(t)\;.
\end{equation}	
	
	This expression could be optimized for the different numerical implementations. Let us assume a case, when the Hamiltonian could be expressed as a following power series:
	
\begin{equation}\label{qe13}
\hat{H}(t) \approx{} \hat{H}(0) + t\hat{H}'(0) + \frac{t^{2}}{2}\hat{H}''(0) + O(t^{3}),
\end{equation}
where the definition of $\hat{H}'(0)$ for arbitrary time-independent function $f$ is following: $\hat{H}'(0)f = [\partial_{t}\hat{H}(t)f]_{t=0}\,$. This gives us an opportunity to express the operator $\partial_{t}\hat{H}(t)$ as follows:

\begin{equation}\label{qe14}
\partial_{t}\hat{H}(t) = \hat{H}'(t) \approx{} \hat{H}'(0) + O(t) \approx{} \frac{\hat{H}(t)-\hat{H}(0)}{t} + O(t).
\end{equation}

	By substituting Eq.(\ref{qe13}) and Eq.(\ref{qe14}) in Eq.(\ref{qe12}), and taking in to account only therms up to second order, we could obtain: 

\begin{equation}\label{qe15}
\hat{P}_{2}(0,t)= 1 + \frac{t}{i\hbar}\cdot{}\frac{\hat{H}(t)+\hat{H}(0)}{2} + \frac{t^{2}}{2(i\hbar)^{2}}\hat{H}^{2}(t)\,.
\end{equation}	
	
	Now we can introduce a new operator $\hat{Z} = \frac{\hat{H}(t)+\hat{H}(0)}{2i\hbar} = \frac{\hat{H}(t)}{i\hbar} + O(t)\,$. Finally, we can write:
	
\begin{equation}\label{qe16}
\hat{P}_{2}(0,t)= 1 + t\hat{Z} + \frac{t^{2}}{2}\hat{Z}^{2}\,.
\end{equation}

	The last expression for the second order propagator could be further transformed in different ways. For example, we could write an operator-exponent form of Eq.(\ref{qe16}) by taking in to account only therms up to second order:

\begin{equation}\label{qe17}
\hat{P}_{2}(0,t)= \exp\big[t\hat{Z}\big]\,.
\end{equation}

	This is a key expression for the \textit{exponential split-operator technique}. However, it could be useful only in the case of separable Hamiltonians. It means that the given operator exponent need to be expressed as a product of some specific operator exponents which could be efficiently calculated. This is quite common for the case of TDSE with some simple electromagnetic interactions under so-called dipole approximation (see for example \cite{JPhB}). Nevertheless, there are some cases when the corresponding operator exponent could not be expressed as a suitable multiplication and therefore can not be efficiently calculated. In this case, it is reasonable to use numerical schemes with a direct calculations of $\hat{Z}$ operator. For example, the optimization of Eq.(\ref{qe16}) in this way could be the following:
	
\begin{equation}\label{qe18}
\hat{P}_{2}(0,t)= \frac{1}{2}\Big(t\hat{Z}+1-i\Big)\Big(t\hat{Z}+1+i\Big)\,.
\end{equation}

	Such multiplication propagator structure could minimize the \textit{RAM} operations with $\psi$-function data in various numerical implementations. 

\newpage

	Finally, we obtain the expression for the \underline{\textit{third-order propagator}} based on Eq.(\ref{qe10}):

\begin{equation}\label{qe19}
\begin{aligned}
\hat{P}_{3}(0,t)= 1 & + \Big[ \frac{t}{i\hbar}\hat{H}(t) - \frac{t^{2}}{2i\hbar}\hat{H}'(t) + \frac{t^{3}}{6i\hbar}\hat{H}''(t) \Big]_{(m=1)} \\
& + \Big[ \frac{t^{2}}{2(i\hbar)^{2}}\hat{H}^{2}(t) - \frac{t^{3}}{3(i\hbar)^{2}}\hat{H}(t)\hat{H}'(t) - \frac{t^{3}}{6(i\hbar)^{2}}\hat{H}'(t)\hat{H}(t) \Big]_{(m=2)} \\
& + \Big[ \frac{t^{3}}{6(i\hbar)^{3}}\hat{H}^{3}(t) \Big]_{(m=3)}\,.
\end{aligned}
\end{equation}

	Here, the $(m)$-number points to the third-order approximation of the corresponding $m$-power therm in Eq.(\ref{qe3}).

\section{Propagation of an electron in focused electromagnetic fields}

	In this section we apply the proposed propagator technique to the calculation of the specific physical problem: propagation of an electron in focused electromagnetic field with vortex electric field component. The main reason of such study is to demonstrate an applicability of our approach for the calculation of time evolution in the case of non-separable Hamiltonian. 
	
\begin{figure*}[ht]
  \centerline{
    \includegraphics{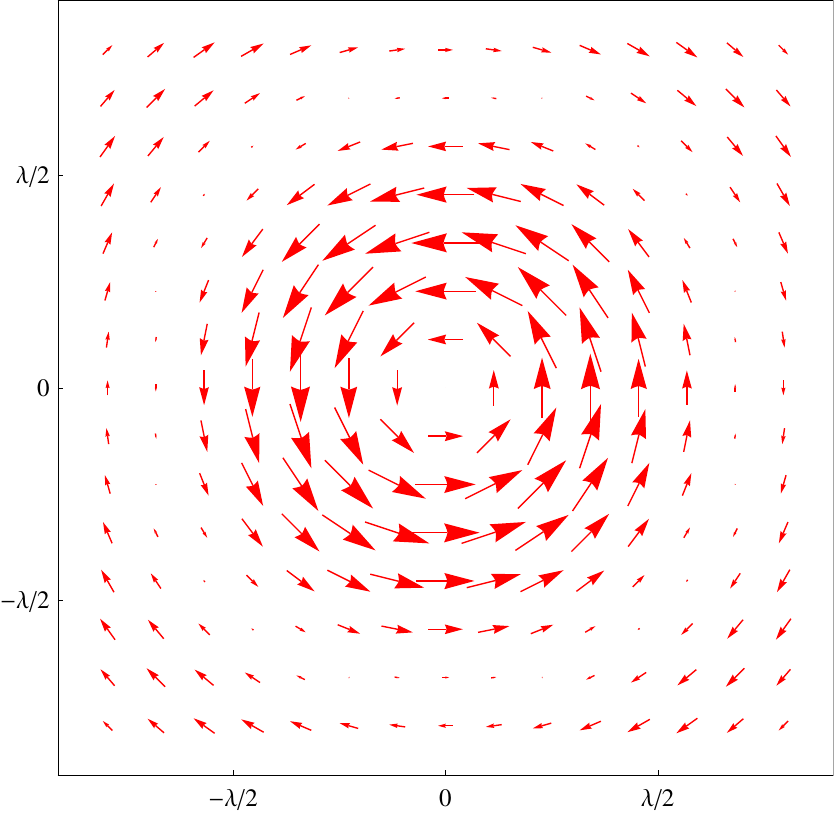}
	}
	\caption{The electric field distribution in \textit{m}-dipole pulse near the focus point (origin). The size of each arrow is proportional to a local field amplitude. The ($xy$) plot corresponds to $z=0$ and the moment of time $t=\pi/2\omega$.}
\label{fig1}
\end{figure*}

	We start with the description of such fields, called \textit{m}-dipole pulses, which correspond to the exact closed-form solution of Maxwell's equations in vacuum and can be generated experimentally, see \cite{DPT} for full details. The expressions for the electric $(\vec{E})$ and magnetic $(\vec{H})$ fields in the most general case of \textit{m}-dipole pulse are the following:

\begin{equation}\label{qe20}
\begin{aligned}
\vec{E}&=-\frac{1}{c}\,\dot{\vec{A}}\,,\\	
\vec{H}&=\nabla\times\vec{A}\,,\\
\vec{A}&=-\nabla\times\vec{Z}\,,\\
\vec{Z}&=\frac{\vec{d}(t+\frac{R}{c})-\vec{d}(t-\frac{R}{c})}{R}\,,
\end{aligned}
\end{equation}
where $R$ is the distance from origin to the point of observation, $c$ is the speed of light in vacuum, $\vec{d}$ is an arbitrary smooth vector function called virtual dipole moment, and the dot symbol above vectors or functions denotes the time partial derivative $\partial_{t}$. After the elementary vector operations, see also \cite{DPT}, we can obtain:

\begin{equation}\label{qe21}
\begin{aligned}
\vec{E}&=\;\vec{n}\times\left[\frac{1}{c^{2}R}\ddot{\vec{d}}_{+}(t,R)+\frac{1}{cR^{2}}\dot{\vec{d}}_{-}(t,R)\right]\,,\\	
\vec{H}&=\;\vec{n}\times\left[\vec{n}\times\frac{1}{c^{2}R}\ddot{\vec{d}}_{-}(t,R)\right]+\frac{1}{cR^{2}}\left[3\vec{n}\Bigr(\vec{n}\cdot\dot{\vec{d}}_{+}(t,R)\Bigr)-\dot{\vec{d}}_{+}(t,R)\right]+\frac{1}{R^{3}}\left[3\vec{n}\Bigr(\vec{n}\cdot{\vec{d}}_{-}(t,R)\Bigr)-{\vec{d}}_{-}(t,R)\right]\,,\\
\vec{A}&=\,-\vec{n}\times\left[\frac{1}{cR}\dot{\vec{d}}_{+}(t,R)+\frac{1}{R^{2}}{\vec{d}}_{-}(t,R)\right]\,,
\end{aligned}
\end{equation}
where $\vec{n}=\vec{R}/R$ is a normal vector, and ${\vec{d}}_{\pm}(t,R)=\vec{d}(t-\frac{R}{c})\pm\vec{d}(t+\frac{R}{c})$. For the calculations below we consider a simple particular case of \textit{m}-dipole pulse which corresponds to: $\vec{d}(\tau)=\vec{\bf{z}}\,d_{0}\sin{(\omega\tau)}$, where $\vec{\bf{z}}$ is a unit vector along $\bf{z}$ axis, $\omega=2\pi{c}/\lambda$ is a characteristic frequency of the radiation, and $d_{0}$ is a constant which could be expressed as a function of average power of incoming radiation ($P$), namely $d_{0}=\sqrt{3}c^{3/2}\sqrt{P}/\omega^{2}$. The vortex structure of the electric field is demonstrated in Fig.\ref{fig1}.

	Next we consider the Schroedinger equation for a non-relativistic electron propagation in \textit{m}-dipole pulse. The Hamiltonian which describes such system is following:

\begin{equation}\label{qe22}
\hat{H}=\frac{1}{2m_{e}}\Big(-i\hbar\,\nabla - \frac{e}{c}\vec{A}\Big)^{2}\,,
\end{equation}
where $\vec{A}$	is determined in Eq.(\ref{qe21}), and $e, m_{e}$ are charge and mass of the electron respectively.
	
	The particular details of the vector potential $\vec{A}$ and corresponding $\vec{d}_{+}$, $\dot{\vec{d}}_{+}$ and $\vec{d}_{-}$ are following:	
\begin{equation}\label{qe23}
\begin{aligned}
\vec{d}_{+}&=\;\vec{\bf{z}}\,d_{0}\big[\sin{(\omega{t}-2\pi{R/\lambda})}+\sin{(\omega{t}+2\pi{R/\lambda})}\big] = 2\vec{\bf{z}}\,d_{0}\sin{(\omega{t})}\,\cos{(2\pi{R/\lambda})}\,,\\	
\dot{\vec{d}}_{+}&=\;2\omega\vec{\bf{z}}\,d_{0}\cos{(\omega{t})}\,\cos{(2\pi{R/\lambda})}\,,\\
\vec{d}_{-}&=\;\vec{\bf{z}}\,d_{0}\big[\sin{(\omega{t}-2\pi{R/\lambda})}-\sin{(\omega{t}+2\pi{R/\lambda})}\big] = -2\vec{\bf{z}}\,d_{0}\cos{(\omega{t})}\,\sin{(2\pi{R/\lambda})}\,,\\
\vec{A}&=\,-\vec{n}\times\left[\frac{1}{cR}\dot{\vec{d}}_{+}(t,R)+\frac{1}{R^{2}}{\vec{d}}_{-}(t,R)\right] = \\
&=\;2d_{0}\,\frac{\vec{\bf{y}}\,x - \vec{\bf{x}}\,y}{R^{3}}\,\cos{(\omega{t})}\,\Bigg[\frac{2\pi{R}}{\lambda}\cos{(2\pi{R/\lambda})}-\sin{(2\pi{R/\lambda})}\Bigg]\,,
\end{aligned}
\end{equation}
where $R=\sqrt{x^{2}+y^{2}+z^{2}}$.

\begin{figure*}[ht]

  \centerline{
    \includegraphics[width = 3cm]{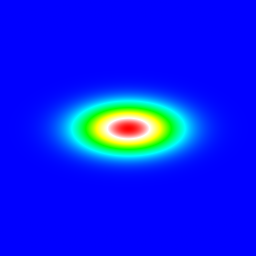}
    \includegraphics[width = 3cm]{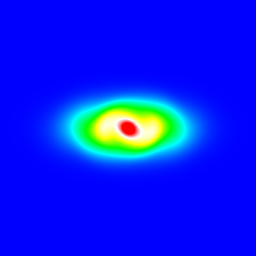}
    \includegraphics[width = 3cm]{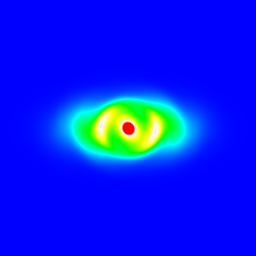}
    \includegraphics[width = 3cm]{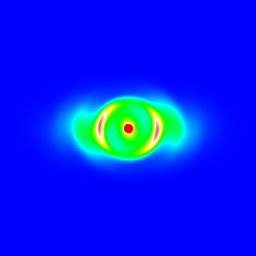}
    \includegraphics[width = 3cm]{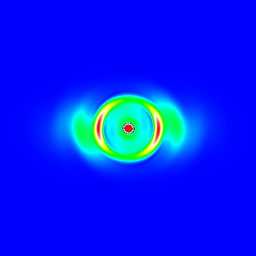}   
	}
	\caption{The snapshots of the electron $\left|\Psi\right|^{2}$ as a function of $x$ and $y$ ($z=0$) at different moments of time. We use linear color scale, where maximum value corresponds to red color and zero corresponds to blue color.}
\label{fig2}
\end{figure*}

	The numerical results of the electron time evolution near the focus of the \textit{m}-dipole pulse are presented in Fig.\ref{fig2}.

	For the numerical calculations of the wave function time evolution (Fig.\ref{fig2}), we used second-order propagator based on Eq.(\ref{qe18}). The initial wave function was specified as Gaussian-form wave packet with zero initial momentum. The calculations were performed in a 3-dimensional grid: $256\times{}256\times{}64$.
	
\newpage

\section{Conclusions}

	In summary, we have obtained the general n-order single-step propagator in closed-form. It could be used for the numerical calculation with any prescribed accuracy of a time evolution in a quantum system of an arbitrary dimensionality and with arbitrary interactions. We have demonstrated its applicability for the case of non-separable Hamiltonian, namely, propagation of an electron in a focused electromagnetic field with vortex electric field component.
	
\section{Acknowledgments}

IG would like to thank Mikhail Emelin and Arkady Gonoskov for valuable discussions.

\end{document}